\documentstyle[twoside,epsf]{article}

\catcode`\@=11
\long\def\@makefntext#1{
\protect\noindent \hbox to 3.2pt {\hskip-.9pt  
$^{{\eightrm\@thefnmark}}$\hfil}#1\hfill}		%CAN BE USED 

\def\@makefnmark{\hbox to 0pt{$^{\@thefnmark}$\hss}}	%ORIGINAL 
        
\def\ps@myheadings{\let\@mkboth\@gobbletwo
\def\@oddhead{\hbox{}
\rightmark\hfil\eightrm\thepage}   
\def\@oddfoot{}\def\@evenhead{\eightrm\thepage\hfil
\leftmark\hbox{}}\def\@evenfoot{}
\def\sectionmark##1{}\def\subsectionmark##1{}}

\oddsidemargin=\evensidemargin
\newcounter{sectionc}\newcounter{subsectionc}\newcounter{subsubsectionc}
\renewcommand{\section}[1] {\vspace{12pt}\addtocounter{sectionc}{1} 
\setcounter{subsectionc}{0}\setcounter{subsubsectionc}{0}\noindent 
        {\tenbf\thesectionc. #1}\par\vspace{5pt}}
\renewcommand{\subsection}[1] {\vspace{12pt}\addtocounter{subsectionc}{1} 
        \setcounter{subsubsectionc}{0}\noindent 
        {\bf\thesectionc.\thesubsectionc. {\kern1pt \bfit #1}}\par\vspace{5pt}}
\renewcommand{\subsubsection}[1] {\vspace{12pt}\addtocounter{subsubsectionc}{1}
        \noindent{\tenrm\thesectionc.\thesubsectionc.\thesubsubsectionc.
        {\kern1pt \tenit #1}}\par\vspace{5pt}}
\newcommand{\nonumsection}[1] {\vspace{12pt}\noindent{\tenbf #1}
        \par\vspace{5pt}}

\newcounter{appendixc}
\newcounter{subappendixc}[appendixc]
\newcounter{subsubappendixc}[subappendixc]
\renewcommand{\thesubappendixc}{\Alph{appendixc}.\arabic{subappendixc}}
\renewcommand{\thesubsubappendixc}
	{\Alph{appendixc}.\arabic{subappendixc}.\arabic{subsubappendixc}}

\renewcommand{\appendix}[1] {\vspace{12pt}
        \refstepcounter{appendixc}
        \setcounter{figure}{0}
        \setcounter{table}{0}
        \setcounter{lemma}{0}
        \setcounter{theorem}{0}
        \setcounter{corollary}{0}
        \setcounter{definition}{0}
        \setcounter{equation}{0}
        \renewcommand{\thefigure}{\Alph{appendixc}.\arabic{figure}}
        \renewcommand{\thetable}{\Alph{appendixc}.\arabic{table}}
        \renewcommand{\theappendixc}{\Alph{appendixc}}
        \renewcommand{\thelemma}{\Alph{appendixc}.\arabic{lemma}}
        \renewcommand{\thetheorem}{\Alph{appendixc}.\arabic{theorem}}
        \renewcommand{\thedefinition}{\Alph{appendixc}.\arabic{definition}}
        \renewcommand{\thecorollary}{\Alph{appendixc}.\arabic{corollary}}
        \renewcommand{\theequation}{\Alph{appendixc}.\arabic{equation}}
        \noindent{\tenbf Appendix \theappendixc #1}\par\vspace{5pt}}
\newcommand{\subappendix}[1] {\vspace{12pt}
        \refstepcounter{subappendixc}
        \noindent{\bf Appendix \thesubappendixc. {\kern1pt \bfit #1}}
	\par\vspace{5pt}}
\newcommand{\subsubappendix}[1] {\vspace{12pt}
        \refstepcounter{subsubappendixc}
        \noindent{\rm Appendix \thesubsubappendixc. {\kern1pt \tenit #1}}
	\par\vspace{5pt}}

\newcommand{\textlineskip}{\baselineskip=13pt}
\newcommand{\smalllineskip}{\baselineskip=10pt}

\def\eightcirc{
\begin{picture}(0,0)
\put(4.4,1.8){\circle{6.5}}
\end{picture}}
\def\eightcopyright{\eightcirc\kern2.7pt\hbox{\eightrm c}} 

\newcommand{\copyrightheading}[1]
        {\vspace*{-2.5cm}\smalllineskip{\flushleft
        {\footnotesize International Journal of Modern Physics C, #1}\\
        {\footnotesize $\eightcopyright$\,\,\, World Scientific Publishing
         Company}\\
         }}

\newcommand{\publisher}[2]{{\begin{center}\footnotesize\smalllineskip 
        Received #1\\
        Revised #2
        \end{center}
        }}

\def\abstracts#1#2#3{{
        \centering{\begin{minipage}{4.5in}\baselineskip=10pt\footnotesize
        \parindent=0pt #1\par
        \parindent=15pt #2\par
        \parindent=15pt #3\par
        \end{minipage}}\par}} 

\newcommand{\bibit}{\nineit}
\newcommand{\bibbf}{\ninebf}
\renewenvironment{thebibliography}[1]
        {\frenchspacing
	 \ninerm\baselineskip=11pt
         \begin{list}{\arabic{enumi}.}
        {\usecounter{enumi}\setlength{\parsep}{0pt}     
         \setlength{\leftmargin 17pt}{\rightmargin 0pt}   %FOR 10--99 ITEMS
         \setlength{\itemsep}{0pt} \settowidth
	{\labelwidth}{#1.}\sloppy}}{\end{list}}

\newcounter{itemlistc}
\newcounter{romanlistc}
\newcounter{alphlistc}
\newcounter{arabiclistc}

\newcommand{\fcaption}[1]{
        \refstepcounter{figure}
	\setbox\@tempboxa = \hbox{\footnotesize Fig.~\thefigure. #1}
	\ifdim \wd\@tempboxa > 5in
           {\begin{center}
	\parbox{5in}{\footnotesize\smalllineskip Fig.~\thefigure. #1}
            \end{center}}
        \else
             {\begin{center}
	     {\footnotesize Fig.~\thefigure. #1}
              \end{center}}
        \fi}

\newcommand{\tcaption}[1]{
        \refstepcounter{table}
	\setbox\@tempboxa = \hbox{\footnotesize Table~\thetable. #1}
        \ifdim \wd\@tempboxa > 5in
           {\begin{center}
         \parbox{5in}{\footnotesize\smalllineskip Table~\thetable. #1}
            \end{center}}
        \else
             {\begin{center}
	     {\footnotesize Table~\thetable. #1}
              \end{center}}
        \fi}

\def\@citex[#1]#2{\if@filesw\immediate\write\@auxout
	{\string\citation{#2}}\fi
\def\@citea{}\@cite{\@for\@citeb:=#2\do
	{\@citea\def\@citea{,}\@ifundefined
	{b@\@citeb}{{\bf ?}\@warning
	{Citation `\@citeb' on page \thepage \space undefined}}
	{\csname b@\@citeb\endcsname}}}{#1}}

\newif\if@cghi
\def\cite{\@cghitrue\@ifnextchar [{\@tempswatrue
	\@citex}{\@tempswafalse\@citex[]}}
\def\citelow{\@cghifalse\@ifnextchar [{\@tempswatrue
	\@citex}{\@tempswafalse\@citex[]}}
\def\@cite#1#2{{$\null^{#1}$\if@tempswa\typeout
	{IJCGA warning: optional citation argument 
	ignored: `#2'} \fi}}

\def\pmb#1{\setbox0=\hbox{#1}
        \kern-.025em\copy0\kern-\wd0
        \kern.05em\copy0\kern-\wd0
        \kern-.025em\raise.0433em\box0}

\def\fnt#1#2{\footnotetext{\kern-.3em
        {$^{\mbox{\scriptsize #1}}$}{#2}}}

\def\fpage#1{\begingroup
\voffset=.3in
\thispagestyle{empty}\begin{table}[b]\centerline{\footnotesize #1}
        \end{table}\endgroup}

\def\runninghead#1#2{\pagestyle{myheadings}
\markboth{{\protect\footnotesize\it{\quad #1}}\hfill}
{\hfill{\protect\footnotesize\it{#2\quad}}}}
\font\tenbf=cmbx10
\font\tenit=cmti10 
\font\tenit=cmti10
\font\bfit=cmbxti10 at 10pt
\font\ninebf=cmbx9
\font\ninerm=cmr9
\font\nineit=cmti9

\font\eightrm=cmr8

\def\lsym{\raise-3pt\hbox{\vbox{\tabskip0pt\offinterlineskip
	\halign{\tabskip0pt plus 1em
	##\tabskip0pt\cr
	$\,\,<\,\,$\cr
	$\,\,\sim\,\,$\cr}}}}
\def\rsym{\raise-3pt\hbox{\vbox{\tabskip0pt\offinterlineskip
     \halign{\tabskip0pt plus 1em
      ##\tabskip0pt\cr
      $\,\,>\,\,$\cr
      $\,\,\sim\,\,$\cr}}}}
\def\qed{\hbox{${\vcenter{\vbox{			%HOLLOW SQUARE
	\hrule height 0.4pt\hbox{\vrule width 0.4pt height 6pt
	\kern5pt\vrule width 0.4pt}\hrule height 0.4pt}}}$}}
\def\theequation{\thesection.\arabic{equation}}		%FOR SETTING Eq.~(1.1)
\begin{document}

\runninghead{S.~J. Mitchell, M.~A. Novotny, Jos\'{e} D. Mu\~{n}oz}
{Application of the Projected Dynamics Method$\cdots$}

\normalsize\textlineskip
\thispagestyle{empty}
\setcounter{page}{1}

\copyrightheading{Vol. 0, No. 0 (1999) 000--000}

\vspace*{0.88truein}

\fpage{1}
\centerline{\bf Application of the Projected Dynamics Method}
\vspace*{0.035truein}
\centerline{\bf to an Anisotropic Heisenberg Model} 
\vspace*{0.37truein}
\centerline{\footnotesize S.~J. Mitchell$^{1,2}$, M.~A. Novotny$^{2}$, Jos\'{e} D. Mu\~{n}oz$^{3,4}$} 
\vspace*{0.015truein}
\centerline{\footnotesize\it $^1$Center for Materials Research and Technology}
\centerline{\footnotesize\it and Department of Physics,}
\centerline{\footnotesize\it Florida State University, Tallahassee, Florida 32306-4351, USA}
\centerline{\footnotesize\it $^2$Supercomputer Computations Research Institute,}
\centerline{\footnotesize\it Florida State University, Tallahassee, Florida 32306-4130, USA}
\centerline{\footnotesize\it $^3$Institute for Computer Applications 1,}
\centerline{\footnotesize\it University of Stuttgart, D-70569 Stuttgart, Germany}
\centerline{\footnotesize\it $^4$Permanent address: Dpto. de F\'isica, Univ. Nacional de Colombia,}
\centerline{\footnotesize\it Bogota D.C., Colombia}
\vspace*{0.225truein}
\publisher{20 August 1999}{20 August 1999}

\vspace*{0.21truein}
\abstracts{
The Projected Dynamics method was originally developed to study metastable decay
in ferromagnetic discrete spin models.
Here, we apply it to a classical, continuous Heisenberg model with anisotropic ferromagnetic
interactions, which evolves under a Monte Carlo dynamic.
The anisotropy is sufficiently large to allow comparison with the Ising model.
We describe the Projected Dynamics method and how to apply it to this continuous-spin system.
We also discuss how to extract metastable lifetimes
and how to extrapolate from small systems to larger systems.
}{}{}

%\vspace*{10pt}
%\keywords{Dynamic Monte Carlo, Heisenberg model, Advanced Algorithms}

%\textlineskip			%) USE THIS MEASUREMENT WHEN THERE IS
%\vspace*{12pt}			%) NO SECTION HEADING

\vspace*{1pt}\textlineskip	%) USE THIS MEASUREMENT WHEN THERE IS
\setcounter{section}{1}
\setcounter{equation}{0}
\section{Introduction}		%) SECTION HEADING
Metastability is a ubiquitous physical phenomenon in which a system remains
near a local free-energy minimum, even though there exists a global minimum with lower
free energy.
If initially near this local free-energy minimum,
the system remains in the vicinity until a long sequence of improbable thermal fluctuations
drive it over a free-energy barrier and into the global free-energy minimum.
The time scale associated with this sequence of improbable events can be very 
long\cite{RikGor,GenMag,Markthis},
and efficient dynamic simulation methods are necessary to conquer the very long time scales \cite{Markthis}.
We discuss here how to extend the Projected Dynamics (PD) method\cite{PDuga,PDprl,PDexact,PDmrs},
which was originally developed for ferromagnetic discrete spin models,
to classical, continuous spin systems like the Heisenberg model with anisotropic ferromagnetic interactions.

The basic concept of the PD method is to project the high-dimensional Markov process
in configuration space,
which describes the metastable decay in microscopic detail,
onto a one-dimensional Markov process in a slow variable.
We choose this variable to be the $z$ component
of the total magnetization, $M_z$.
For the ferromagnetic Ising model, or the ferromagnetic Heisenberg model 
with sufficient anisotropy in the $z$ direction,
below the critical temperature,
the free energy projected onto $M_z$
is symmetric about $M_z=0$ in zero applied field, and it displays two equal minima.
When a magnetic field $\vec{H}=H_z \hat{z}$ is applied, 
this symmetry is broken.
The global free-energy minimum is the stable well, 
in which the majority of the spins are aligned parallel with the external field,
while the local minimum is the metastable well,
in which the majority of the spins are anti-parallel with the field.

To study the time scale for the system to escape from the metastable well,
the system is prepared with all spins aligned in the $+\hat{z}$ direction for times $t<0$.
This initial state corresponds to the stable well for $H_z =+\infty$.
At time $t=0$, the field is instantaneously reversed to a finite negative value,
and the system is evolved under a Monte Carlo dynamic.
The system quickly relaxes to the bottom of the metastable well,
in the vicinity of which it remains until thermal fluctuations drive it over the saddle
point and into the stable well.
We define the escape time $\tau$ as the time (in Monte Carlo steps per spin, MCSS) 
required for the system to first reach $M_z=0$.
We define the lifetime of the metastable state as $\langle \tau \rangle$,
the escape time averaged over many statistically independent escapes.

For the Ising model, or the Heisenberg model with sufficiently large $z$ anisotropy,
in two or three dimensions,
the dominant mechanism to escape the metastable well at small fields is by nucleation and growth 
of droplets of the stable phase within a background of the metastable 
phase\cite{RikGor,RTMS94,Chui2,Chui1,Nowak2,Nowak1}.
If $R$ is the radius of a droplet of the stable phase,
we can define a critical droplet radius $R_c$.
Subcritical droplets, $R<R_c$, are dominated by the tendency 
to minimize surface free energy and tend to shrink.
Super-critical droplets, $R>R_c$, are dominated by the tendency to 
minimize bulk free energy and tend to grow.
For critical droplets, $R=R_c$, the probabilities to grow and shrink are equal.
We also define\cite{RTMS94} the typical separation between critical droplets as $R_0$.

For the Ising model, four different droplet regimes have been found\cite{RTMS94}
as the strength of the external field is varied for a system of fixed linear size $L$.
In the coexistence regime, $L<R_c$, the system is not large enough to contain a complete critical droplet.
In the single droplet (SD) regime, $R_0\gg L\gg R_c$, only one droplet of the stable phase 
nucleates and grows to take over the system.
In the multi-droplet (MD) regime, $R_0\ll L$, many droplets nucleate and grow to take over the system.
In the strong-field (SF) regime, $R_c \approx 1$ spin, the droplet picture is no longer useful.

Figure 1 shows $\langle \tau \rangle$ for the anisotropic kinetic Heisenberg model studied
in this paper (see Sec. 2 for definition) with $L=16$ as a function of $1/|H_z|$.
The SD, MD, and SF regimes are labeled.
Because of the long time scales,
we were unable to simulate at very weak fields where the coexistence regime is expected.
When plotted in this way, the data in the SD and MD regimes should asymptotically fall on 
straight lines.
Neglecting prefactors in the nucleation rate,
the Kolmogorov-Johnson-Mehl-Avrami (KJMA) theory\cite{RikGor,GenMag,RTMS94,Ramos99} of independent droplet nucleation and growth
predicts that the ratio of the slopes in the SD and MD regimes should be $d+1$,
where $d=2$ is the spatial dimension of the spin lattice.
The difference between our measured ratio of the slopes of 2.5 and the expected
value of 3 is the result of non-exponential prefactors in the nucleation rate\cite{RTMS94}.
The decay is characterized by widely different time scales in the three regimes.
The PD method provides a way to understand and conquer these disparate time scales.

\begin{figure}
%Fig.~1
\vspace{0.5in}
%\centerline{\vbox{\hrule width5cm height0.001pt}}
\begin{center}
{\epsfxsize=3in \epsfysize=2in \epsfbox{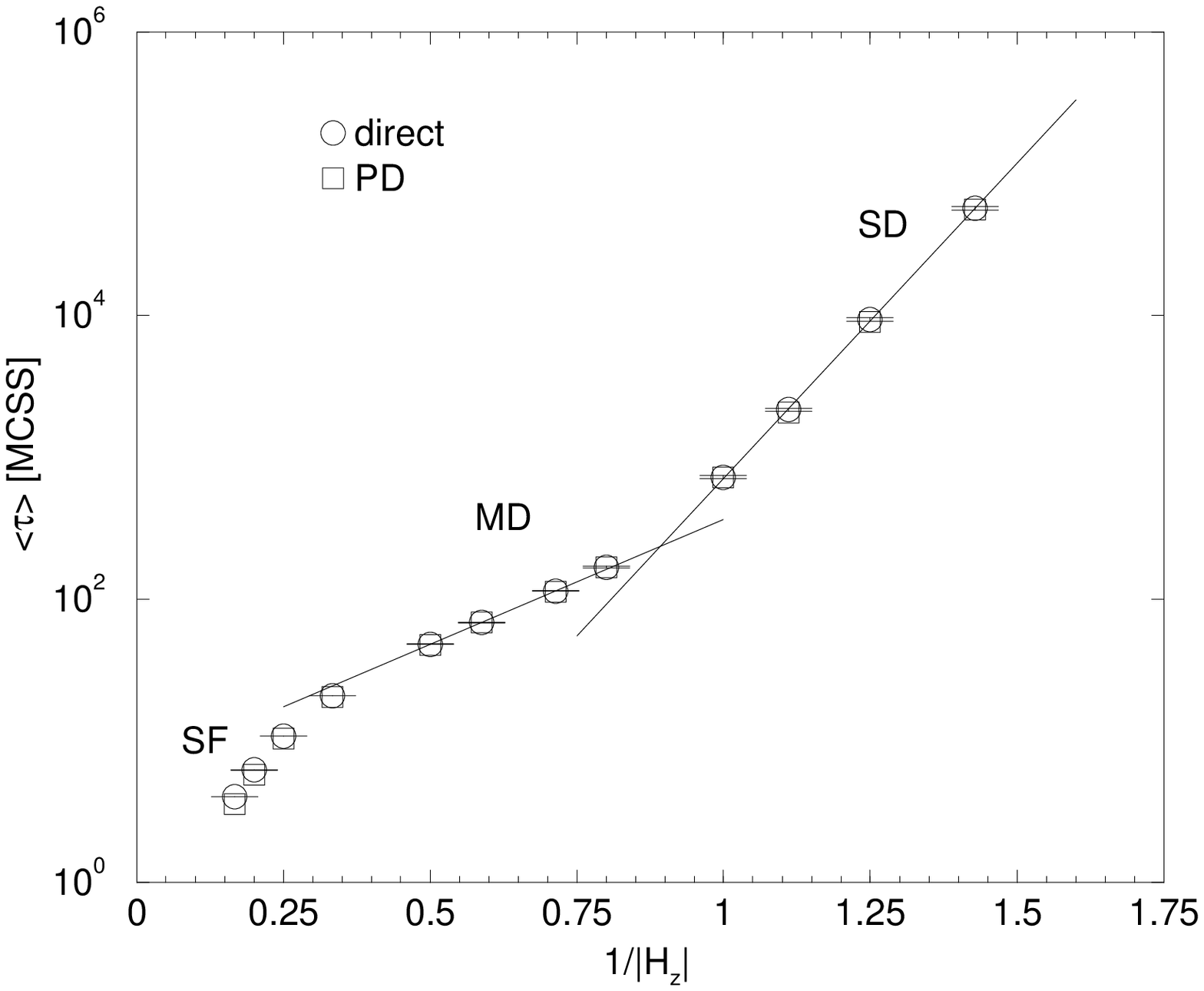}}
\end{center}
%\centerline{\vbox{\hrule width5cm height0.001pt}}
\vspace*{10pt}
\fcaption{
The metastable lifetime for $J_x=J_y=1$, $J_z=2$, $T=1$, and $L=16$.
The circles indicate the lifetimes measured directly by simulating 1000 escapes.
The squares indicate the lifetimes measured by the PD method
with 1000 escapes.
The error bars indicate the statistical error in only the direct measurement of the lifetime (circles).
The ratio of the slopes of the two lines is approximately 2.5.
The strong field (SF), multi-droplet (MD), and single droplet (SD) regimes are labeled.
}
\end{figure}

This paper is organized in the following way.
Section 2 introduces the ferromagnetic Heisenberg model with anisotropic interactions
and the Monte Carlo dynamic.
Section 3 describes the PD method.
Section 4 discusses how to extrapolate to larger system sizes.
Section 5 gives a summary of our results and a brief discussion of
possibilities for future work.
\bigskip

\section{Model}
We consider a two-dimensional, square array of three-dimensional, classical,
Heisenberg (XYZ) spins with anisotropic ferromagnetic interactions.
The Hamiltonian is given by
\begin{equation}
{\cal H} = -\sum_{\langle i j \rangle} (J_x s_{i x} s_{j x} +J_y s_{i y} s_{j y} +J_z s_{i z} s_{j z})
- \vec{H} \cdot \sum_i \vec{s}_i,
\end{equation}
where $\vec{s}_i$ refers to the $i^{th}$ Heisenberg spin,
$\sum_i$ is a sum over all spins,
$\vec{H}$ is the magnetic field,
$J_x$, $J_y$, and $J_z$ are the interactions in the $x$, $y$, and $z$ directions respectively,
and $\sum_{\langle i j\rangle}$ is a sum over all nearest-neighbor pairs.
For all systems studied here, we choose $J_x=J_y=1$, $J_z=2$, 
and $\vec{H}=H_z \hat{z}$ where $\hat{z}$ is perpendicular 
to the plane containing the spins and $H_z<0$.
All systems studied were $L \times L$ squares with $L=16$ or $32$ and periodic boundary conditions.

The system is evolved with a Monte Carlo dynamic after the field reversal discussed in the
previous section.
There is a unique choice of deterministic microcanonical spin dynamic\cite{Landauthis}.
However, since any dynamic satisfying detailed balance will 
give the correct equilibrium distribution,
there is no unique choice of Monte Carlo dynamic\cite{Chui2,Chui1,Nowak2,Nowak1}
In this work, we use the following Monte Carlo dynamic.
First, choose a spin location $i$ at random.
Call $\vec{s}_{old}$ the current orientation of spin $i$.
Next, randomly choose a new trial orientation $\vec{s}_{new}$ for spin $i$,
uniformly distributed over the solid angle
by choosing the polar angle $\phi$ uniformly on the interval $[0,2 \pi)$
and choosing $\cos(\theta)$ uniformly on $[-1,+1]$.
The trial spin $\vec{s}_{new}$ is either accepted or rejected as the next orientation for
spin $i$ according to the Glauber acceptance probability
\begin{equation}
P(\vec{s}_{new}|\vec{s}_{old})=\{1+\exp{[\beta (E_{new}-E_{old})]}\}^{-1},
\end{equation}
where $\beta=(k_B T)^{-1}$
and $E_{new}$ and $E_{old}$ are the energies of the system with the new
and old spin orientations respectively.
The temperature is represented by $T$,
and $k_B=1$ is Boltzmann's constant.
In these simulations, we use $T=1$, which is below the critical temperature $T_c$.
Using the maximum of the specific heat and the susceptibility of the $z$ component of the magnetization,
we found that for the anisotropy used here, $T_c\approx 1.75$.
\bigskip

\setcounter{section}{3}
\setcounter{equation}{0}
\section{Projected Dynamics Method}
As shown in Fig. 1, metastable lifetimes can be extremely long.
Thus, simulations of metastable decay require fast algorithms\cite{Markthis}.
However, since we are studying the dynamic behavior of the system under
the Monte Carlo dynamic defined above, 
we cannot speed up the simulation by techniques that alter the dynamic.
Thus conventional fast equilibrium algorithms, like cluster updates\cite{Swendsen,Wolff,Chayes}, 
are not acceptable, and a different approach is needed.

The dynamic behavior of the system is a random walk in the configuration space
of $2 L^2$ continuous degrees of freedom.
The essence of the PD method is to project this Markov process
onto a Markov process in the one-dimensional $M_z$ space.
It is important to point out that the actual walk through $M_z$ space is not Markovian.
There are memory terms, 
but neglecting these memory terms appears to be a very reasonable approximation.
To define the random walk in $M_z$ we should measure the transition probabilities per
Monte Carlo step (mcs) between each $M_z$ state and every other $M_z$ state.
However, since there is a continuum of $M_z$ states for the Heisenberg model,
this is not possible.
Therefore, we must lump $M_z$ states together into discrete, coarse-grained states, or bins,
and measure the transition probabilities between these bins.
In the Ising model, for which the PD method was originally developed\cite{PDuga},
the $M_z$ states are discrete, and no coarse-graining is required.

To simplify the projected Markov process,
we further impose that the bins be large enough that only
transitions between neighboring bins are possible.
The largest possible change in $M_z$ that can occur during one mcs is $\pm 2$,
corresponding to $\vec{s}_{i}=-\hat{z} \rightarrow \vec{s}_{i}=+\hat{z}$
or vice versa.
Thus, we divide the $M_z$ space into $L^2$ equal bins of width 2.
We define an index or bin number for the lumped states as $n={\rm Int}[(L^2-M_z)/2]$,
such that the $n=0$ bin includes the states $M_z \in [-L^2,-L^2+2)$
and the $n=L^2-1$ bin includes the states $M_z \in [L^2-2,L^2)$.
The $n=L^2$ bin contains only the $M_z=L^2$ state.
With this choice of binning, only transitions from bin $n$ to bins $n+1$
and $n-1$ are possible in one mcs.
We define the growth probability $g(n)/L^2$ as the probability of undergoing 
a transition from $n$ to $n+1$ in one mcs.
Analogously, the shrinkage probability $s(n)/L^2$ is the probability of undergoing
a transition from $n$ to $n-1$ in one mcs.
We define the escape time as the time to first enter the bin $n=n_{\rm stop}$.
For an escape time cut-off of $M_z=0$, $n_{\rm stop}=L^2/2$.

Before discussing how to measure the growth and shrinkage probabilities,
it is useful to return to our discussion of droplet theory.
For the fields and anisotropy studied here,
most spins are either approximately aligned or approximately anti-aligned with the magnetic field.
The number of spins nearly aligned with the field approximates the number of stable spins in the system.
When a droplet of the stable phase nucleates,
the width of the domain wall is on the order of one lattice spacing.
To a rough approximation (exact for Ising)\cite{PDexact}, the $M_z$ bin index $n$
gives the number of stable spins in the system,
such that $n=0$ corresponds to all spins in the metastable phase
and $n=n_{\rm stop}=L^2/2$ corresponds to about $1/2$ of the spins in the metastable phase.
The growth, $g(n)/L^2$, and shrinkage, $s(n)/L^2$, probabilities are then the probabilities 
that the volume of the stable phase will grow or shrink respectively.

There are two aspects to measuring the growth and shrinkage probabilities.
First, we must generate a series of sample configurations for each lumped $M_z$ state.
Second, we must measure the growth and shrinkage probabilities for each sample configuration.
The simplest and most straight forward way to generate the sample configurations
is simply to perform dynamic simulations of many escapes.
The growth and shrinkage probabilities are then obtained by integrating over all the spins in the 
configuration and averaging over configurations.

There are several ways to obtain these probabilities for a single configuration.
Since we have an analytic expression for the probability density function for the orientations 
of a single spin (the Glauber probability with a normalization constant),
one might be inclined to analytically integrate this distribution function
to find the probability that a single spin will move the configuration into the next bin
after one mcs.
The Broad Histogram method\cite{BH1,BHWang,BHXY,BHHeisenberg} uses a similar analytic analysis tool 
for measuring the density of states in equilibrium.
Unfortunately, we have been unable to analytically perform the necessary double integrations.
These integrals could be calculated numerically by the trapezoidal method
or by some other numerical method,
but because these integrals must be calculated for every spin each mcs,
numerical methods are too computationally intensive for practical use in this case.
There are also too many parameters to construct look-up tables of reasonable size.
Therefore, we use a very simple counting method, i.e.\ a Monte Carlo integration
over spins and configurations.

To calculate the growth and shrinkage probabilities by the counting method,
we simply perform the dynamic Monte Carlo simulation and keep track 
of the bin number of the current configuration.
We record the number of mcs spent in bin $n$ ($H_n$),
as well as the number of times the system jumped from bin $n$ into bins $n+1$ ($G_n$) and $n-1$ ($S_n$).
The Monte Carlo estimates for the growth and shrinkage probabilities are then
$g(n)/L^2=G_n/H_n$ and $s(n)/L^2=S_n/H_n$.
It is important to note that all methods for obtaining the growth and shrinkage
probabilities should give identical results in the limit of an infinite number of samples.

Figure 2 shows the growth and shrinkage probabilities for $H_z=-0.9$ and $L=16$, obtained from 1000 escapes
with $n_{\rm stop}=237$.
The most obvious features are the three crossings where the growth and shrinkage probabilities
are equal.
Each crossing corresponds to an extremum in the projected free-energy landscape\cite{PDuga,PDmrs}.
From left to right, the crossings give the locations of the bottom of the metastable well, the saddle point
and the bottom of the stable well.
The location of the saddle point is extremely difficult to find by any other method
but is given quite simply by the PD method.
The noise in the growth and shrinkage probabilities could be reduced with better statistics,
but this data is adequate to obtain very reasonable values of the average lifetime $\langle \tau \rangle$.

%Fig.~2
\begin{figure}
\vspace{0.5in}
%\centerline{\vbox{\hrule width5cm height0.001pt}}
\begin{center}
{\epsfxsize=3in \epsfysize=2in \epsfbox{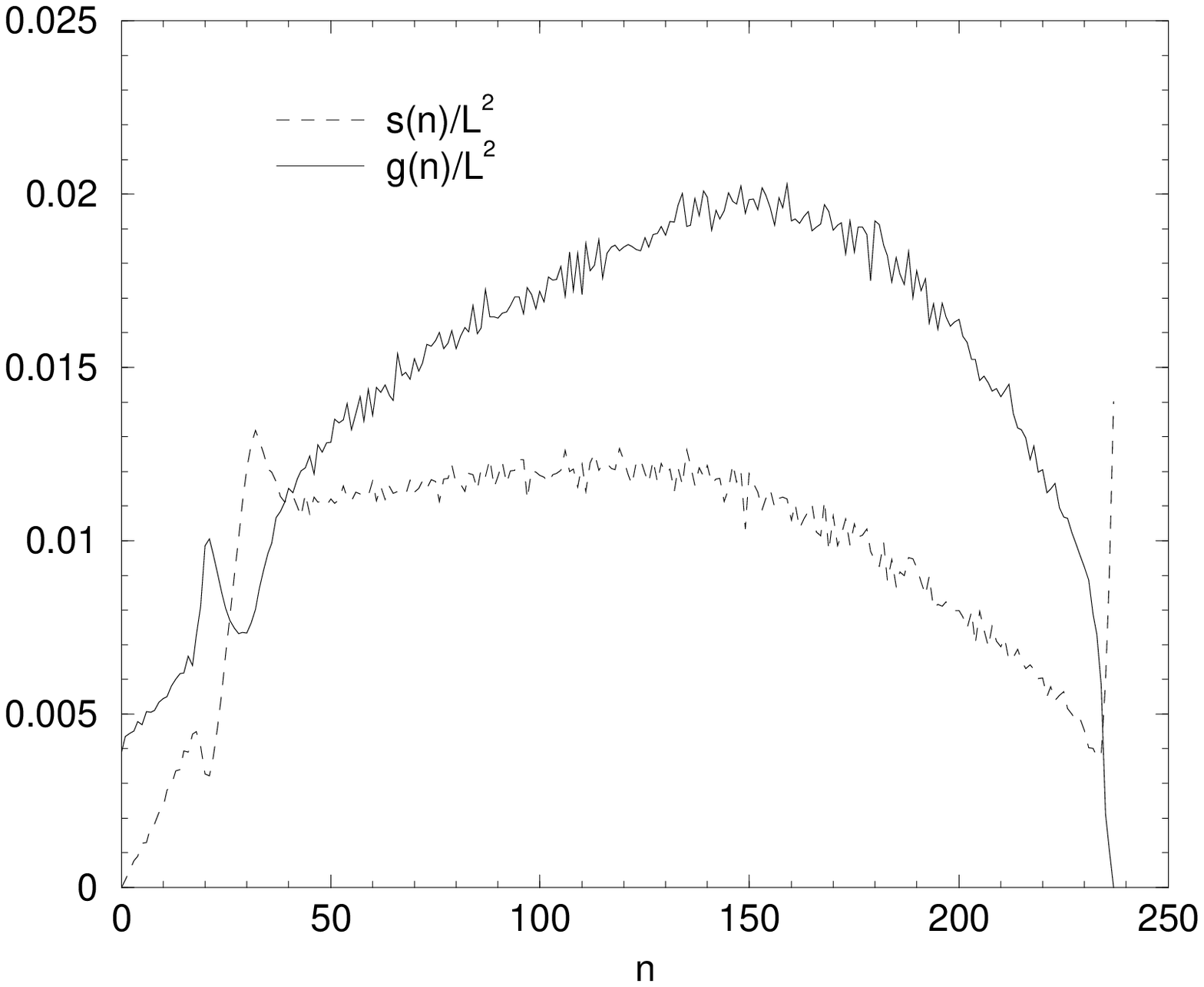}}
\end{center}
%\centerline{\vbox{\hrule width5cm height0.001pt}}
\vspace*{10pt}
\fcaption{
The growth (solid line) and shrinkage (dashed line) probabilities 
for $J_x=J_y=1$, $J_z=2$, $T=1$ and $L=16$ at $H_z=-0.9$
measured from 1000 escapes with $n_{\rm stop}=237$.
The three crossings from left to right correspond to the bottom of the metastable well,
the saddle point, and the bottom of the stable well.
}
\end{figure}

To compute the lifetime from the growth and shrinkage probabilities,
we first compute the residence time $h(n)$, which is the average number of MCSS
that the system spends in bin $n$ per escape.
The average lifetime can be found with the following recursive relations\cite{PDuga,PDprl,PDexact}:
\begin{eqnarray}
h(n_{\rm stop}-1) & = & \frac{1}{g(n_{\rm stop}-1)},\nonumber \\
h(n) & = & \frac{1+s(n+1) h(n+1)}{g(n)},\\
\langle \tau \rangle & = & \sum_{n=0}^{n_{\rm stop}-1} h(n).\nonumber
\end{eqnarray}
Higher moments of $\tau$ can be found in a similar way\cite{PDuga,Lee}.

Figure 1 shows a comparison of $\langle \tau \rangle$ measured directly 
by averaging $\tau$ over 1000 escapes with 
the measurement from the growth and shrinkage probabilities.
Although the growth and shrinkage probabilities give useful information,
simply measuring them is no faster than measuring 
$\langle \tau \rangle$ by direct simulation with the same number of escapes.
However, measurements of the average lifetime can be performed more quickly
if we extrapolate the growth and shrinkage
probabilities to larger system sizes.
\bigskip

\section{Size Extrapolation}
Given $g(V,n)$ and $s(V,n)$ for a system of $V=L^2$ spins,
we can extrapolate to find $g(2 V,n)$ and $s(2 V,n)$.
The following relations\cite{PDuga} are used to extrapolate from $V \rightarrow 2 V$:
\begin{equation}
g(2 V,n) \approx \frac{\sum_{i=0}^n h(V,n-i) h(V,i) [g(V,n-i)+g(V,i)]}{\sum_{i=0}^n h(V,n-i) h(V,i)},
\end{equation}
\begin{equation}
s(2 V,n) \approx \frac{\sum_{i=0}^n h(V,n-i) h(V,i) [s(V,n-i)+s(V,i)]}{\sum_{i=0}^n h(V,n-i) h(V,i)},
\end{equation}
where $n \in [0,V/2]$ insures that the late-time approach to the stable phase 
does not alter the extrapolated probabilities.
These relationships can be applied recursively for even larger systems.

Figure 3 shows the growth and shrinkage probabilities 
for $L=32$ and $H_z=-0.9$, obtained by direct simulation of 1000 escapes for the $L=32$ system
and by extrapolating twice from $L=16$.
For $L=16$, $n_{\rm stop}=L^2/2=128$.
However, the extrapolated growth and shrinkage probabilities 
are only reasonable for $n\le V/2$ for the smallest $V$ used in the extrapolation.
Thus, we must use $n_{\rm stop}=128$ for $L=32$, even though this does not correspond to $M_z=0$
for the $L=32$ system.
With the cut-off bin $n_{\rm stop}=128$, $\langle \tau \rangle = 277$~MCSS 
from the extrapolated probabilities
and $\langle \tau \rangle = 295 \pm 15$~MCSS 
from the direct measurement of the $32^2$ lattice also with $n_{\rm stop}=128$.
Using this extrapolation we have saved a factor of about four in simulation time.

Additional simulation time may be saved in computing $h(n)$ for different stopping bins.
Since $h(n)$ depends on $n_{\rm stop}$,
without knowledge of $g(n)$ and $s(n)$, a new simulation must be performed to calculate $h(n)$
for different $n_{\rm stop}$.
However, $g(n)$ and $s(n)$ are independent of $n_{\rm stop}$, within statistical error,
and therefore, once $g(n)$ and $s(n)$ have been obtained for a large value of $n_{\rm stop}$,
$h(n)$ is easily calculated for any smaller cut-off.

\begin{figure}
%Fig.~3
\vspace{0.5in}
%\centerline{\vbox{\hrule width5cm height0.001pt}}
\begin{center}
{\epsfxsize=3in \epsfysize=2in \epsfbox{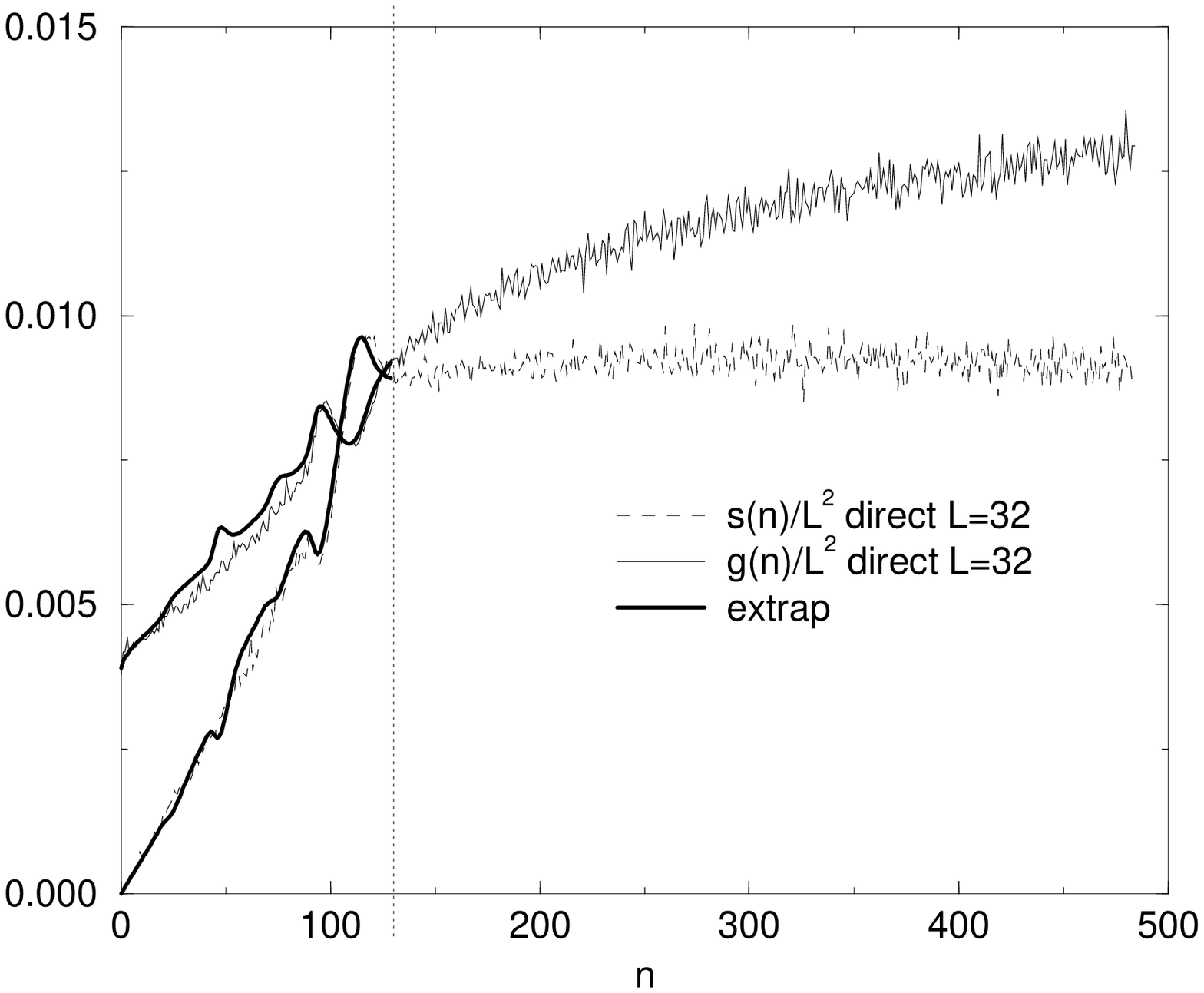}}
\end{center}
%\centerline{\vbox{\hrule width5cm height0.001pt}}
\vspace*{10pt}
\fcaption{
The growth (solid line) and shrinkage (dashed line) probabilities 
for $J_x=J_y=1$, $J_z=2$, $T=1$ and $L=32$ at $H_z=-0.9$
measured directly from simulating 1000 escapes.
The heavy solid lines indicate the growth and shrinkage probabilities
for $L=32$, obtained by extrapolating twice from $L=16$, which only extends to bin $n=128$,
denoted by the vertical dotted line.
}
\end{figure}
\bigskip

\section{Summary}
In summary, we have shown that the droplet picture\cite{RikGor,GenMag,RTMS94,Chui2,Chui1,Nowak2,Nowak1}
provides a reasonable description for the Heisenberg model
with the anisotropic ferromagnetic interactions used here.
We have also shown that the metastable escape is well described by 
the projected Markov process in the one-dimensional $M_z$ space.
We have applied the PD method over a wide range of fields and
obtained very good estimates of the average lifetimes.
Extrapolations to larger system sizes have also been shown to be accurate.
Such extrapolations reduce the amount of computer time required to perform the simulations.

It was found for discrete spin models\cite{Markthis,PDprl,PDexact,PDmrs} that 
forcing the system to escape from the metastable well by inserting a wall moving in the $M_z$
space, gave accurate estimates of the average lifetime and allowed speed-ups of several 
orders of magnitude in computer time.
By forcing the system to escape, lifetimes could be measured at weaker fields (longer lifetimes)
than were previously accessible.
Future work on the Heisenberg model will attempt to extend the range of fields accessible in 
a reasonable amount of computer time by application of assisted escape methods
similar to the forced escaped methods used for the discrete models.

Still another way to save computational time in metastable decay of discrete spin models
is by extrapolation to weak fields, and hence to long lifetimes\cite{PDuga}.
Utilizing this technique for our continuous model
would require us to obtain the functional form for the single spin flip growth and shrinkage
probabilities, requiring that we are able to perform the Broad Histogram\cite{BH1,BHWang,BHXY,BHHeisenberg}
type double integrations described in Sec. 3.

\bigskip

\nonumsection{Acknowledgment}
The authors would like to thank 
G. Brown,
G. Korniss
and
P.~A. Rikvold
for useful discussions.  
Supported in part by the NSF through grant number DMR-9871455,
the Supercomputer Computations Research Institute which is
funded by the U.S.\ Department of Energy and the State of Florida, 
the Center for Materials Research and Technology,
and the National Energy Research Scientific Computing Center.
\bigskip

\nonumsection{References}

\end{document}